\documentclass[12pt]{article}
\usepackage{authblk}
\usepackage{lipsum}

\usepackage{comment}
\usepackage{pdfpages}
\usepackage{sidecap}
\usepackage{subcaption}
\usepackage{setspace}
\usepackage[ruled,linesnumbered]{algorithm2e}
\usepackage{xcolor}
\usepackage{graphicx,natbib,amsfonts,amsthm,amssymb}
\usepackage[colorlinks,citecolor=black,urlcolor=black,linkcolor = black]{hyperref}
\usepackage{amsmath,mathrsfs}
\usepackage{wrapfig,lipsum,booktabs}
\usepackage{appendix}
\usepackage{enumitem}
\usepackage{multirow}

\usepackage{xr}

\def\spacingset#1{\renewcommand{\baselinestretch}%
{#1}\small\normalsize} \spacingset{1}

\renewcommand{\baselinestretch}{1.3}

\newtheorem{theorem}{Theorem}

\newtheorem{definition}{Definition}
\newtheorem{example}{Example}

\newtheorem{simulation}{Simulation}

\renewcommand{\textit}{}

\newcommand{\E}{{\rm I}\kern-0.18em{\rm E}}

\renewcommand{\hat}{\widehat}
\newtheorem{prop}{Proposition}

 \newcommand{\p}{{\rm I}\kern-0.18em{\rm P}}
 \newcommand{\1}{{\rm 1}\kern-0.24em{\rm I}}

\newcommand{\R}{{\rm I}\kern-0.18em{\rm R}}

\newcommand{\rom}[1]{\uppercase\expandafter{\romannumeral #1\relax}}
\newcommand{\ee}{\end{equation*}}
\newcommand{\ben}{\begin{equation}}
\newcommand{\een}{\end{equation}}
\newcommand{\bmln}{\begin{multline*}}
\newcommand{\emln}{\end{multline*}}

\def\l{\left}
\def\r{\right}
\newcommand{\mc}{\mathcal}

\usepackage[english]{babel}
\usepackage{blindtext}

\usepackage{xpatch}

\renewcommand{\thefootnote}{\arabic{footnote}}

\allowdisplaybreaks

\usepackage{JASA_manu}
\makeatletter
\newcommand\blfootnote[1]{%
  \begingroup
  \renewcommand{\@makefntext}[1]{\noindent\makebox[1.8em][r]#1}
  \renewcommand\thefootnote{}\footnote{#1}%
  \addtocounter{footnote}{-1}%
  \endgroup
}
\makeatother

\title{\Large \textsc{\bf{Neyman-Pearson and equal opportunity: when economic efficiency meets societal fairness in classification}}}

\date{} 

\begin{document}





\author[1]{Jianqing Fan}
\author[2]{Xin Tong}
\author[3]{Yanhui Wu}
\author[4] {Lucy Xia}
\author[5]{Shunan Yao}
\affil[1]{\footnotesize Department of Operations Research and Financial Engineering, Princeton University;}
\affil[2]{\footnotesize Department of Data Sciences and Operations, University of Southern California;}
\affil[2,3]{\footnotesize Faculty of Business and Economics, The University of Hong Kong;}
\affil[4]{\footnotesize Department of ISOM, HKUST Business School, Hong Kong University of Science and Technology;} 
\affil[5]{\footnotesize Department of Mathematics, Hong Kong Baptist University}

\maketitle

\blfootnote{The authors are listed alphabetically. Correspondence jqfan@princeton.edu and xint@hku.hk.}


\begin{abstract}


Organizations have increasingly used statistical algorithms to make decisions. While fairness and other social considerations are important in these automated decisions, economic efficiency remains crucial to the survival and success of organizations. Therefore, a balanced focus on both algorithmic fairness and economic efficiency is essential for promoting fairness in real-world data science solutions.  Among the first efforts towards this dual focus, we incorporate the equal opportunity (EO) constraint into the Neyman-Pearson (NP) classification paradigm. Under this new NP-EO framework, we derive the oracle classifier, propose finite-sample-based classifiers that satisfy population-level fairness and efficiency constraints with high probability, and demonstrate the statistical and social effectiveness of our algorithms on simulated and real datasets.

\textbf{keyword}: classification, economic efficiency, algorithmic fairness, Neyman-Pearson, equal opportunity.
\end{abstract}


\section{Introduction}

Recently, the U.S. Department of Justice and the Equal Employment Opportunity Commission warned employers that used artificial intelligence to hire workers for potential unlawful racial discrimination.\textsuperscript{\footnotemark{}}\footnotetext{\ ``AI Hiring Tools Can Violate Disability Protections, Government Warns," Wall Street Journal, May 12, 2022. https://www.wsj.com/articles/ai-hiring-tools-can-violate-disability-protections-government-warns-11652390318} Earlier, Amazon was accused of gender bias against women in its deployment of machine learning algorithms to search for top talents.\textsuperscript{\footnotemark{}}\footnotetext{\ ``Amazon scraps secret AI recruiting tool that showed
bias against women," October 11, 2018. https://www.reuters.com/article/us-amazon-com-jobs-automation-insight-idUSKCN1MK08G}

\setlength{\belowdisplayskip}{5pt} \setlength{\belowdisplayshortskip}{5pt}
\setlength{\abovedisplayskip}{5pt} \setlength{\abovedisplayshortskip}{5pt}

Evidence that algorithmic decision-making exhibits systematic bias against certain disadvantageous social groups has been accumulating in labor markets \citep{chalfin2016productivity, lambrecht2019algorithmic} and also growing in many other areas, including credit lending, policing, court decisions, and healthcare treatment \citep{arnold2018racial, kleinberg2018human, bartlett2022consumer, obermeyer2019dissecting, fuster2022predictably}. To address the public concern of algorithmic fairness, several studies propose to regulate algorithmic design such that disadvantageous groups must receive non-disparate treatments \citep{barocas2016big, kleinberg2017inherent, corbett-davies2017algorithmic, barocas-hardt-narayanan}. Statistically, this means that, in carrying out its predictive task, an algorithm ought to prioritize the fairness-related construction, such as purposefully equalizing certain error types of concern. However, efficiency loss could occur as these fairness-related designs may limit the prediction accuracy \citep{kleinberg2017inherent}.

Along this line of research, we study algorithmic design when organizations aim to achieve interlocking objectives under fairness constraints. Consider that a bank uses an algorithmic classifier to decide whether to approve a loan application based on default status prediction. If fairness is a primary social concern, the disparity between denial rates of qualified applicants by non-credit attributes, such as gender or race, will not be tolerated. The bank, however, is more concerned about economic efficiency, resulting in a potential conflict with social fairness. How to resolve this conflict in terms of the trade-off between societal fairness and classification accuracy has been the major focus of existing studies \citep{corbett-davies2017algorithmic,valdivia2021fair,chzhen2022minimax,celis2019classification,menon2018cost,zeng2022bayes,zeng2024bayes}.

The key innovation of our study lies in moving beyond the conventional focus of classification accuracy. We do not equate classification accuracy directly with economic efficiency. Instead, we frame efficiency in the context of an organization’s specific goals, such as profitability and business stability. In the aforementioned example of credit lending, we can decouple classification accuracy into two parts — the type I error (i.e., the probability of misclassifying a default case as non-default) and the type II error (i.e., the probability of misclassifying a non-default case as default). When financial security is paramount, the bank will prioritize controlling the type I error over the type II error. To cope with such real-world challenges, we propose a novel classification framework that simultaneously addresses both efficiency and fairness objectives while explicitly incorporating asymmetric priorities in efficiency considerations.

The \emph{efficiency} part of our framework is based on the Neyman-Pearson (NP) classification paradigm  \citep{cannon2002learning,scott2005neyman}. This paradigm controls the type I error\footnote{It is worth noting that the practical meaning of type I error depends on how classes $0$ and $1$ are defined. To avoid confusion, we refrain from using the terms false positive and false negative in the development of the algorithm, as the prioritized class $0$ could represent either the positive or negative class depending on the context. Moreover, in many classification problems (e.g., dogs vs. cats), there is no connotation of positive and negative classes at all. } (i.e., the probability of misclassifying a 0 instance as 1) under some desired level $\alpha$ (referred to as the NP constraint) while minimizing the type II error (i.e., the probability of misclassifying a 1 instance as 0). In the loan application example, we label default status as class 0, as misclassifying a default case is more financially consequential than mis-flagging a non-default case. The asymmetric treatment of the NP paradigm permits flexible control over the more consequential error type. 

In the literature, error asymmetry in classification is commonly addressed by cost-sensitive learning \citep{Elkan01,ZadLanAbe03,menon2018cost}, which reformulates the objective function by assigning different weights to errors. While this approach offers significant merits and practical value, determining appropriate costs for misclassification errors can be challenging due to the lack of established standards across applications and the risk of misspecification. For this reason, we adopt the NP paradigm as an alternative choice - one that has already been implemented in several loan companies. Although the selection of the type I error bound $\alpha$ is still required, it can be guided by policy considerations \citep{equal1979,menon2018cost} and thus is more practical and interpretable in real-world implementation.\footnote{Federal Reserve Bank of St. Louis: https://fred.stlouisfed.org/series/DRCCLACBS}  Further discussions are included in the Supplementary Material Section B.4.

The \emph{fairness} part of our framework adapts a relaxed version of the equality of opportunity (EO) constraint proposed by \citep{hardt2016equality}. Assuming class 1 is the favored outcome, the EO constraint requires achieving the same type II error in all sensitive groups (e.g., race or gender); in the context of loan applications, this means that denial rates of qualified applicants should be equalized in different groups. The relaxation we adopt eases the exact rate-equality requirement by allowing a pre-specified $\varepsilon$ difference \citep{donini2018empirical, agarwal2018reductions}. For discussion purposes, we retain the term `EO constraint' to describe this relaxed formulation. While we have chosen EO as our fairness metric in this study, our framework can be extended to incorporate other classic fairness measures, such as Demographic Parity and Equalized Odds\footnote{We refer interested readers to \cite{caton2024fairness} for a broader overview of fairness measures.}, as will be discussed at the end of this work. For now, we focus on EO to conduct a detailed and comprehensive study.

Integrating the efficiency and fairness components outlined above, we propose the novel NP-EO paradigm: for any given $\alpha, \varepsilon \in (0,1)$ that control the probability of type I error and EO respectively (see \eqref{eq:NP_EO}), are the NP constraint for economic efficiency and the EO constraint for social fairness simultaneously feasible? We provide a positive answer to this question. Moreover, leveraging the generalized Neyman-Pearson Lemma, we derive an NP-EO oracle classifier.  Guided by this oracle, we construct finite-sample-based classifiers that respect the population-level NP and EO constraints with high probability. The form of the oracle inspires us to take an umbrella algorithm perspective; that is, we wish to adjust the commonly used methods (e.g, logistic regression, random forest, gradient boosting tree, neural nets) to the NP-EO  paradigm in a universal way and propose a provable algorithm for this overarching goal. Similar to the original NP umbrella algorithm developed in \cite{tong2018neyman} and its variant for corrupted labels in \cite{Yao.Rava.Tong.James.2022}, we employ an order statistics approach and do not require distributional assumptions on data in algorithmic development. However, the technicalities here are much more involved than in the NP umbrella algorithm because we need to determine two thresholds (instead of one) simultaneously.  
In simulation studies, we demonstrate that NP-EO classifiers are the only classifiers that guarantee both NP and EO constraints with high probability. This advantage of the NP-EO classifiers is further demonstrated in a case study concerning credit card approvals.

This paper contributes to the emerging literature on algorithmic fairness. Existing studies have focused on algorithmic bias due to data sampling and engineering \citep{rambachan2019bias,cowgill2020algorithmic}, the construction of fairness conditions \citep{hardt2016equality, kleinberg2017inherent}, and the way of incorporating ethical concerns into algorithmic optimization \citep{corbett-davies2017algorithmic}, among others.  Methods for ensuring fairness in predictive algorithms can be broadly categorized into three groups: (1) pre-processing, where algorithms are trained on debiased data \citep{zemel2013learning,lum2016statistical,zeng2024bayes}, (2) in-processing, where fairness constraints are integrated during the training process \citep{goh2016satisfying,zeng2024bayes}, and (3) post-processing, where fairness is enforced after models are trained \citep{celis2019classification, chen2023post,chzhen2019leveraging,denis2024fairness,gaucher2023fair,menon2018cost,zeng2022bayes,zeng2024bayes}. Our work falls into the post-processing category. In this category, \cite{celis2019classification} introduced linear fractional constraints, while \cite{chen2023post} proposed a novel bias score to accommodate diverse group fairness measures. \cite{chzhen2019leveraging} and \cite{denis2024fairness} explored the use of both labeled and unlabeled data to improve prediction accuracy. Furthermore, \cite{gaucher2023fair} established a connection between regression and classification under demographic parity constraints, extending their findings to encompass all linear fractional performance measures. \cite{menon2018cost} derived oracle classifiers and corresponding estimators within the cost-sensitive learning framework.

{Most relevant to our study are the works of \cite{zeng2024bayes} and \cite{li2023fairee}.} \cite{zeng2024bayes} explored Bayes optimal classifiers using the Generalized Neyman-Pearson Lemma. Their work spans pre-, in-, and post-processing steps while developing oracle classifiers under various fairness metrics. However, our work departs from theirs in the following two important aspects. First, the oracle classifiers derived in \cite{zeng2024bayes} primarily focus on maximizing overall accuracy. They use a cost-sensitive (CS) framework to address differing priorities for the two error types, whereas our approach constrains the population-level type I error. Although NP and CS are equivalent at the population level with a one-to-one correspondence, knowing a type I error bound does not reveal exact costs. When justifying specific costs is difficult, bounding type I error provides more intuitive and policy-relevant guidance. {Second, to construct sample-based classifiers, \cite{zeng2024bayes} proposes a plug-in rule, FPIR, which solves an estimated one-dimensional fairness equation at a specified disparity level but lacks population-level theoretical guarantees.} In contrast, our method constructs the threshold non-parametrically using order statistics, which allows us to derive explicit oracle bounds on the population-level type I error and the type II error disparity, with high-probability guarantees. These theoretical results are not available in \cite{zeng2024bayes}. {\cite{li2023fairee} also uses order statistics to construct the threshold in their fair classifier; however, it requires more stringent conditions for high-probability bounds and does not address type I error control.}

Generally speaking, we build on existing works that focus on the trade-off between fairness and accuracy to address the more fundamental social science problem: the trade-off between economic efficiency and social equality. This complex challenge can be concretized in a three-way trade-off — between type I error, type II error, and fairness constraints — that lies at the core of the NP-EO framework. 
 Some researchers advocate a social-planning approach, in which the algorithmic designer models a social welfare function that captures an explicit preference for a certain socially desirable objective \citep{kleinberg2018human, rambachan2020economic}. While this approach provides a useful benchmark to evaluate social welfare in the presence of ethical considerations, how to put it into practice is a great challenge. Social preferences are often difficult to measure and have to be approximated by some measurable outcomes. These proxies can be mis-measured and lead the predictive outcomes astray, as demonstrated in \cite{mullainathan2017does} and \cite{obermeyer2019dissecting}.

Alternative to the social-planning approach, our approach is from a regulatory perspective, in which a decision maker can pursue their objective after obeying a certain regulatory constraint. Existing algorithmic designs under the regulatory framework \citep{corbett-davies2017algorithmic} do not explicitly cope with the efficiency-equality trade-off. Regulatory failure is likely to occur when the efficiency loss caused by the fairness constraint is significant. Our proposed NP-EO approach provides a framework to detect algorithmic bias, evaluate the social loss caused by self-interested algorithms, and regulate algorithms to maintain the regulatory goal while permitting users sufficient freedom to achieve economic efficiency.

In the algorithmic fairness literature, many criteria were proposed to define ``fairness''; see \cite{barocas-hardt-narayanan} and references within.  Our work does not intend to introduce another new fairness criterion.  Rather, our framework is flexible enough that the EO constraint can potentially be replaced by other well-defined fairness criteria, and the NP constraint can also be replaced by other efficiency priorities.  Such efficiency-fairness dual constraints have the potential to be implemented as long as their population versions are simultaneously feasible.

The rest of the paper is organized as follows. Mathematical settings of the Neyman-Pearson equal opportunity (NP-EO) paradigm are introduced in Section \ref{sec:np-eo}. Then, Section \ref{sec:np-eo classifier} presents the NP-EO oracle classifier. We introduce two NP-EO umbrella algorithms and provide theoretical justification in Section \ref{sec:method}. Numerical studies are presented in Section \ref{sec:num}. Finally, we conclude with a discussion. Lemmas, proofs, and other technical materials are relegated to the Supplementary Material. 

\section{Neyman-Pearson equal opportunity (NP-EO) paradigm}\label{sec:np-eo}
\subsection{Mathematical setting and preliminaries}

Let $(X, S, Y)$ be a random triplet where $X \in \mc X \subset \R^d$ represents $d$ features, $S$ denotes a sensitive attribute that takes values from $\{a, b\}$, and $Y$ denotes the class label that takes values from $\{0,1\}$. Not every feature in $X$ needs to be \emph{neutral};  we partition the features into $X$ and $S$ to emphasize that we will specifically consider a classifier's societal impacts related to $S$.   We denote by $\p$ a generic probability measure whose meaning will be clear in context, and denote respectively by $\p_Z$ and $\p_{\mc B}$ the probabilities taken with respect to the randomness of $Z$ and $\mc B$, for any random variable $Z$ and random set $\mc B$.  Let $\phi: \mc X \times \{a,b\} \mapsto \{0,1\}$ be a classifier. The (population-level) type I error and type II error of $\phi$ are defined as 
\begin{align*}
    R_0(\phi) := \p\l(\phi(X, S) \neq Y \mid Y = 0\r)\quad \text{and }\quad 
    R_1(\phi) := \p\l(\phi(X, S) \neq Y \mid Y = 1\r)\,,
\end{align*}
respectively. Next, we denote the type I/II error conditional on the sensitive attribute by $R_y^s (\phi) := \p\l(\phi(X, S) \neq Y \mid Y=y, S = s\r)$ for $y \in \{0, 1\}$ and $s \in \{a, b\}$.  Then it follows that, 
\begin{align}\label{eq: error decomposition}
    R_y (\phi) = \p (\phi(X,S) \not = Y | Y = y )= R_y^a(\phi)\cdot p_{a|y} + R_y^b(\phi)\cdot p_{b|y}\,,
\end{align}
where $p_{s|y} = \p(S = s \mid Y = y)$ for $s\in\{a,b\}$. Each $p_{s|y}$ is assumed to be non-zero, and we use $X^{y,s}$ as a shorthand of $X \mid \{Y = y, S = s\}$ for $y \in \{0, 1\}$ and $s \in \{a, b\}$. 
Throughout the paper, we consider class 1 as the `favored' outcome for \emph{individuals}, such as `being hired', `receiving promotion', `admission to a college', or `non-default', and class 0 as the less-favored outcome for \emph{individuals}.  In the meantime, we understand class 0 as the class that \emph{organizations} are concerned about and try to avoid, such as `default'.

\subsection{Equality of opportunity (EO)}

Let $L_y(\phi) := \l|R_y^a(\phi) - R_y^b(\phi)\r|$. In the literature of algorithmic fairness, a popular notion of fairness, coined as `equalized odds' (or `separation'), requires absolute equality across social groups for any outcome, or $L_0(\phi) = L_1(\phi) = 0$ in our notation; see \cite{barocas-hardt-narayanan} and the references therein. \cite{hardt2016equality} formulated a less-stringent condition, referred to as `equality of opportunity', which only requires $L_1(\phi) = 0$. That is, qualified people from different social groups have equal opportunities to obtain the `favored' outcome. This weaker notion of fairness is consistent with the advocacy of productive equity in social science and is acceptable in a wide range of social contexts. 


The requirement of absolute equality is, however, not practical for finite-sample-based classifiers: due to the randomness of data, the population-level condition $L_1(\phi) = 0$ can hardly be achieved from any finite-sample training procedure. Thus, researchers (e.g., \cite{donini2018empirical, agarwal2018reductions})  worked on a relaxed criterion:
\begin{align}\label{eqn: eo constraint}
    L_1(\phi) \leq \varepsilon\,,
\end{align}
for some pre-specified small $\varepsilon$. This condition states that equality of opportunity is satisfied if, for two groups, the difference in the probabilities of falsely classifying a ``favored'' outcome as ``unfavored'' is sufficiently small. This less stringent criterion offers a flexible level of tolerance and could be achieved by finite sample procedures with high probability. In this paper, we adopt the relaxed EO condition described by equation \eqref{eqn: eo constraint} and refer to it as the EO constraint. Furthermore, we refer to $L_1(\phi)$ as the type II error disparity of $\phi$.

\subsection{Neyman-Pearson (NP) paradigm}

Like other fairness criteria, the EO constraint draws a boundary to incorporate the societal concern of fairness in algorithmic decision-making. In the fairness literature, it was combined with some general loss functions (e.g., \cite{woodworth2017learning}). For example, it was incorporated into the \emph{classical} classification paradigm, which minimizes the overall classification error, i.e., a weighted average of type I and type II errors, with the weights equal to the marginal probabilities of the two classes. In many applications, however, these weights do not reflect the relative importance of different error types; as a consequence, classifiers under the classical paradigm could have undesirably high type I or type II errors. The inclusion of a fairness criterion can further complicate the problem with an (unintended) redistribution of the two types of classification errors, as will be shown by Example \ref{ex:simple} in Section~\ref{sec:np-eo classifier}.  


Recall the loan application example. A bank wishes to classify loan applicants to control the default risk (controlling the type I error) and gain ample business opportunities (maximizing $1 -$  type II error). The problem is that the two types of errors are statistically in conflict, and the bank has to balance the trade-off between the goals. 
Regulation from fairness concerns (e.g., through the EO constraint) may help lift the bank's bias against certain social groups and enlarge its business opportunities (lower type II error), but it could also expose the bank to greater default risk (higher type I error). 

To cope with the above problem, we propose using the Neyman-Pearson (NP) paradigm \citep{cannon2002learning,scott2005neyman, rigollet2011neyman}, which solves:
\begin{align}\label{eqn: NP_oracle}
    \min_{\phi : R_0(\phi) \leq \alpha}R_1(\phi)\,,
\end{align}
where $\alpha \in (0, 1)$ is a user-specified constant. 
In the loan example, an NP oracle classifier would control the risk of classifying a default applicant as a non-default one, helping banks manage financial risk; after securing financial safety, it minimizes the chances of
misclassifying a non-default applicant, giving banks the maximum possible business opportunities.


%



\subsection{NP-EO paradigm}

We propose the NP-EO paradigm as follows:
\begin{align}\label{eq:NP_EO}
    \min_{R_0(\phi) \leq \alpha, L_1(\phi) \leq \varepsilon} R_1(\phi)\,,
\end{align}
where $\alpha, \varepsilon\in(0,1)$ are pre-specified numbers.  Program \eqref{eq:NP_EO} has joint constraints: the NP constraint $R_0(\phi)\leq \alpha$ which ensures the most important part of economic efficiency, and the EO constraint $L_1(\phi) \leq \varepsilon$ which enforces the social fairness restriction. 
In this arrangement, the direct impact of the EO constraint on the type I error $R_0$ is isolated, and the conflict between efficiency and equality is absorbed by the type II error $R_1$, which is assumed to be economically less consequential. On the population level, we will derive an NP-EO oracle classifier, i.e.,  a solution to program \eqref{eq:NP_EO}. On the sample level, we will construct finite sample-based classifiers that respect the two constraints in \eqref{eq:NP_EO} with high probability.

Returning to the loan application example, a bank is concerned with two private goals---controlling the default risk ($R_0$) and expanding business opportunity ($R_1$)---and a social goal of maintaining equal opportunity (a small $\lvert R_1^a - R_1^b \rvert$). With the NP-EO paradigm, the risk-control goal is achieved by the constraint  $R_0(\phi) \leq \alpha$,  where $\alpha$ is a risk level chosen by the bank, and the social goal is achieved by the constraint $L_1(\phi) \leq \varepsilon$, where $\varepsilon$ is determined by regulation or social norms. With these two goals, the bank has to be modest in the business expansion goal --- potentially paying the cost of having a larger chance of misclassifying non-defaulters as defaulters. While this cost could be more significant for startup banks, it is small for established banks that have a large customer base.

\section{NP-EO oracle classifier}\label{sec:np-eo classifier}

In this section, we establish an NP-EO oracle classifier, a solution to the constrained optimization program \eqref{eq:NP_EO}. 
The establishment of an NP-EO oracle classifier demands efforts because (i) the simultaneous feasibility of the NP and EO constraints is not clear on the surface, and (ii) the functional form of the oracle is unknown. 


Let $f_{y, s}(\cdot)$ be the density function of $X^{y,s}$  and  
$
    F_{y,s}(z) = \p\l(f_{1,s}(X) \leq z f_{0,s}(X) \mid Y = y, S = s\r)
$, for each $y \in \{0, 1\}$ and $s \in \{a, b\}$. Moreover, we denote, for any $c_a, c_b$,
\begin{align}\label{eq:oracle_classifier}
    \phi^{\#}_{c_a, c_b}(X, S) &= \1\{f_{1,a}(X) > c_a f_{0,a}(X)\}\1\{S = a\}  
    + \1\{f_{1,b}(X) > c_b f_{0,b}(X)\}\1\{S = b\}\,.
\end{align}

\begin{theorem}\label{thm:oracle}
For each $y \in \{0,1\}$ and $s \in \{a,b\}$, we assume (i) $f_{y,s}$ exists, (ii) $F_{y,s}(z)$ is continuous on $[0,\infty)$, and (iii) $F_{y,s}(0) = 0$ and $\lim_{z \rightarrow \infty} F_{y,s}(z) = 1$. Then there exist two non-negative constants $c_a^*$ and $c_b^*$ such that $\phi^\#_{c^*_a, c^*_b}$ is an NP-EO oracle classifier. \footnote{ Assumption (i) guarantees the existence of density functions for each subclass. Assumptions (ii) and (iii) ensure that for any $y \in \{0, 1\}$ and $s \in \{a, b\} $, the distribution function $ F_y^s$ is continuous and spans \( (0,1) \). These assumptions are standard and weak, analogous to those used in the Neyman-Pearson lemma for determining the most powerful test.} 
\end{theorem}


The solution is intuitive:  within each class, the choice should be a likelihood ratio, and two different thresholds must be used to satisfy two constraints.
The proof of Theorem \ref{thm:oracle} is relegated to the Supplementary Materials. Here, we briefly sketch the idea. The existence assumption of $f_{y,s}$'s is necessary to write down a classifier in the form of equation \eqref{eq:oracle_classifier}. The assumptions on $F_{0,a}$ and $F_{0,b}$ ensure that $R_0^a$ and $R_0^b$ can take any value in $(0, 1)$ by varying thresholds $(c_a, c_b)$. Therefore, $R_0$, as a convex combination of $R_0^a$ and $R_0^b$, can achieve an arbitrary level $\alpha\in(0,1)$.  Similarly, the conditions $F_{1,a}$ and $F_{1,b}$ guarantee that $R_1^a$ and $R_1^b$ can take any value in $(0,1)$. Thus, $L_1 = \varepsilon$ can be achieved for arbitrary $\varepsilon\in(0,1)$. In sum, the conditions in Theorem \ref{thm:oracle} easily ensure that proper choices of thresholds are sufficient to satisfy either NP or EO constraint. The reasoning for simultaneous feasibility is involved, and we will demonstrate it in a special case shortly.  

Note the Neyman-Pearson lemma implies that the NP oracle classifier (i.e., a solution to program \eqref{eqn: NP_oracle}) is of the form
\begin{align*}
    \phi(x, s) = \1\l\{\frac{f_{1,s}(x)\cdot p_{s|1}}{f_{0,s}(x)\cdot p_{s|0}} > c\r\}= 
    \1\l\{\frac{f_{1,a}(x)}{f_{0,a}(x)} > c\frac{p_{a\mid0}}{p_{a\mid1}}\r\}\cdot \1\{s=a\} + \1\l\{\frac{f_{1,b}(x)}{f_{0,b}(x)} > c\frac{p_{b\mid0}}{p_{b\mid1}}\r\}\cdot \1\{s=b\}\,,
\end{align*}
for some constant $c$ such that the NP constraint takes the boundary condition. It is easy to see that the last expression in the above display is of the form in equation \eqref{eq:oracle_classifier}. If the NP oracle classifier satisfies the EO constraint, then it is also an NP-EO oracle. {\color{black}
If the NP oracle classifier fails to satisfy the EO constraint, e.g., $R_1^a(\phi) - R_1^b(\phi) > \varepsilon$, the generalized Neyman-Pearson lemma (Theorem S.3 in Supplementary Materials) indicates that it suffices to find a classifier of the form in equation \eqref{eq:oracle_classifier} with non-negative $c_a$ and $c_b$ that achieves $R_0 = \alpha$ and $R_1^a(\phi) - R_1^b(\phi) = \varepsilon$. To see this, note that part (iii) of Theorem S.3 in Supplementary Materials states that such a classifier, given its existence, minimizes $R_1(\phi)$ among all classifiers that satisfy $R_0 \leq \alpha$ and $R_1^a(\phi) - R_1^b(\phi) \leq \varepsilon$, and thus constitutes an NP-EO oracle classifier. Thus, it remains to find the two non-negative values $c_a$ and $c_b$.
}


\begin{SCfigure}[][h]
     \centering
     \includegraphics[width = 0.7\textwidth, height = 7cm]{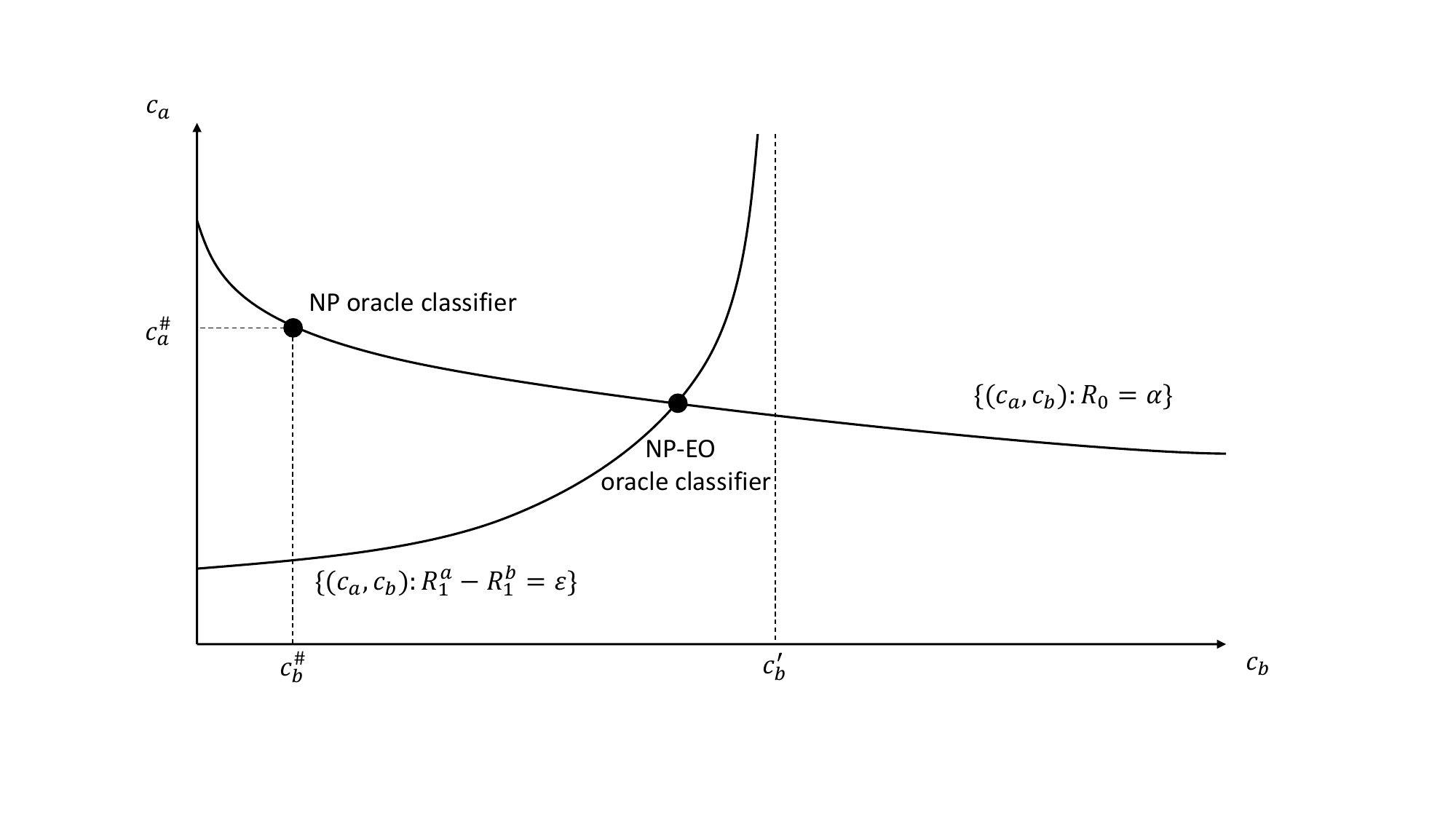}
     \vspace{-3em}
    \caption{\footnotesize Feasibility of NP-EO oracle.  The downward curve represents the critical values $c_a$ and $c_b$ in the classifier \eqref{eq:oracle_classifier} such that the type I error $R_0$ is $\alpha$, whereas the upward curve depicts the classifiers satisfying $R_1^a - R_1^b = \varepsilon$.  The intersection of these two curves gives the critical values for the NP-EO classifier. }\label{fig:npeo_existence}
   \end{SCfigure}


The existence of such a pair in one scenario is illustrated by Figure \ref{fig:npeo_existence}, where we assume that $R^a_1-R^b_1 > \varepsilon$ for the NP oracle. More general discussion can be found in the proof of Theorem \ref{thm:oracle}. In Figure \ref{fig:npeo_existence}, the vertical and horizontal axes are $c_a$ and $c_b$,  representing respectively the $S = a$ and $S= b$ part of the thresholds in the classifier in  \eqref{eq:oracle_classifier}. Thus, every point in the first quadrant represents such a classifier. In this figure, $c_b'$ is the constant such that its corresponding $R_1^b = 1 - \varepsilon$. The solid downward curve represents pairs $(c_a, c_b)$ such that $R_0 = \alpha$; note that $R_0(\phi^\#_{c_a, c_b}) = (1 - F_{0,a}(c_a))\cdot p_{a|0} + (1 - F_{0,b}(c_b))\cdot p_{b|0}$ so when $R_0$ is fixed at $\alpha$, $c_a$ is non-increasing as $c_b$ increases, which is shown in Figure \ref{fig:npeo_existence}. At the same time, the solid upward curve represents the threshold pairs $(c_a, c_b)$ such that $R_1^a - R_1^b = \varepsilon$.  Since
$
    R_1^a(\phi^\#_{c_a, c_b}) - R_1^b(\phi^\#_{c_a, c_b}) = F_{1,a}(c_a) - F_{1,b}(c_b),
$
so when $R_1^a - R_1^b$ is fixed at $\varepsilon$, $c_a$ is non-decreasing when $c_b$ increases, and hence the curve should be upward. 
As indicated in Figure \ref{fig:npeo_existence}, it can be shown that there must be an intersection of the two curves, which satisfies both the NP and EO constraints. Then, the generalized Neyman-Pearson lemma implies that the intersection must be an NP-EO oracle classifier.  Nevertheless, the uniqueness of the solution is not necessarily guaranteed if the function $F_{y,s}(z) = \p\l(f_{1,s}(X) \leq z f_{0,s}(X) \mid Y = y, S = s\r)$ has ``flat'' regions for some $y$ and $s$, as this could result in multiple pairs of $(c^*_a, c^*_b)$ yielding the same values for $R_0$, $R_1$, and $L_1$.

We recognize that this non-uniqueness does not affect the design of our algorithm. Both $\text{NP-EO}_{\text{OP}}$ and $\text{NP-EO}_{\text{MP}}$ utilize the umbrella approach, which means they employ any scoring-type learning algorithm as a base and use order statistics to construct potential thresholds. The pair that minimizes the empirical type II error, while ensuring high-probability NP and EO control, is then selected. This construction is unaffected by the potential non-uniqueness of the oracle.

Now we rationalize results in Theorem \ref{thm:oracle} on an intuitive level. Theorem \ref{thm:oracle} states that an NP-EO oracle comprises two parts, namely, $S = a$ component and $S = b$ component. This is understandable because, as long as a classifier $\phi$ takes into consideration the protected attribute $S$, it can always be rewritten as a two-part form, i.e.,
$
    \phi(X, S) = \phi^a(X)\cdot\1\{S = a\} + \phi^b(X)\cdot\1\{S = b\},
$
where $\phi^a(\cdot) = \phi(\cdot, a)$ and $\phi^b(\cdot) = \phi(\cdot, b)$. Then, given the two-part form, it is not surprising that the best $\phi^a$ and $\phi^b$, in terms of group-wise type II error performance for a type I error level,  adopt density ratios as scoring functions. Thus, as long as the two thresholds are adjusted so that NP and EO constraints are satisfied, the classifier in the form of equation \eqref{eq:oracle_classifier} will have smaller $R_1^a$ and $R_1^b$ than other feasible classifiers and thus a smaller $R_1$. We now present a simple example to illustrate the NP-EO oracle. 

\begin{example}\label{ex:simple}
Let $X^{0,a}, X^{1,a}, X^{0,b}$ and $X^{1,b}$ be $\mc N(0, 1), \mc N(4, 1), \mc N(0, 9)$ and $\mc N(4, 9)$ distributed random variables, respectively, and set $\p(S = a, Y = 0) = \p(S = a, Y = 1) = \p(S = b, Y =1) = \p(S=b, Y = 1) = 0.25$. Then, the Bayes classifier is $\phi_{\text{Bayes}} = \1\{X > 2\}$ and the NP oracle classifier for $\alpha =0.1$ is $\phi_{\text{NP}} = \1\{X > 2.56\}$.\textsuperscript{\footnotemark{}}\footnotetext{\ In this example, the sensitive attribute $S$ does not appear in the Bayes classifier or in the NP oracle classifier because the thresholds are the same for the $S = a$ and $S = b$ components. Thus, $S$ can be omitted due to the specific setup of this model.} If $\alpha=\varepsilon=0.1$, the NP-EO oracle classifier is $\phi_{\text{NP-EO}} = \1\{X > 3.20\}\1\{S=a\} + \1\{X > 2.53\}\1\{S=b\}$. The graphical illustration of this example is depicted in Figure \ref{fig:npeo_example}. We can calculate that $R_0(\phi_{\text{Bayes}}) = 0.138$, $R_1(\phi_{\text{Bayes}}) = 0.138$ and $L_1(\phi_{\text{Bayes}}) = 0.230$, violating both NP and EO constraints. The NP oracle, compared with the Bayes classifier, has a larger threshold. Consequently, $R_0(\phi_{\text{NP}}) = 0.1$, $R_1(\phi_{\text{NP}}) = 0.195$ and $L_1(\phi_{\text{NP}}) = 0.241$. The NP oracle classifier satisfies the NP constraint but violates the EO constraint. The NP-EO oracle classifier is more subtle. Its $S=a$ part threshold is larger than that of NP oracle classifier whereas the $S=b$ part threshold is slightly smaller,  resulting in $R_0(\phi_{\text{NP-EO}}) = 0.100$, $R_1(\phi_{\text{NP-EO}}) = 0.262$ and $L_1(\phi_{\text{NP-EO}}) = 0.1$ so that the NP-EO oracle classifier satisfies both NP and EO constraints. 

\begin{figure}[t]
\begin{center}
    \includegraphics[width = 0.7\textwidth, height = 8cm]{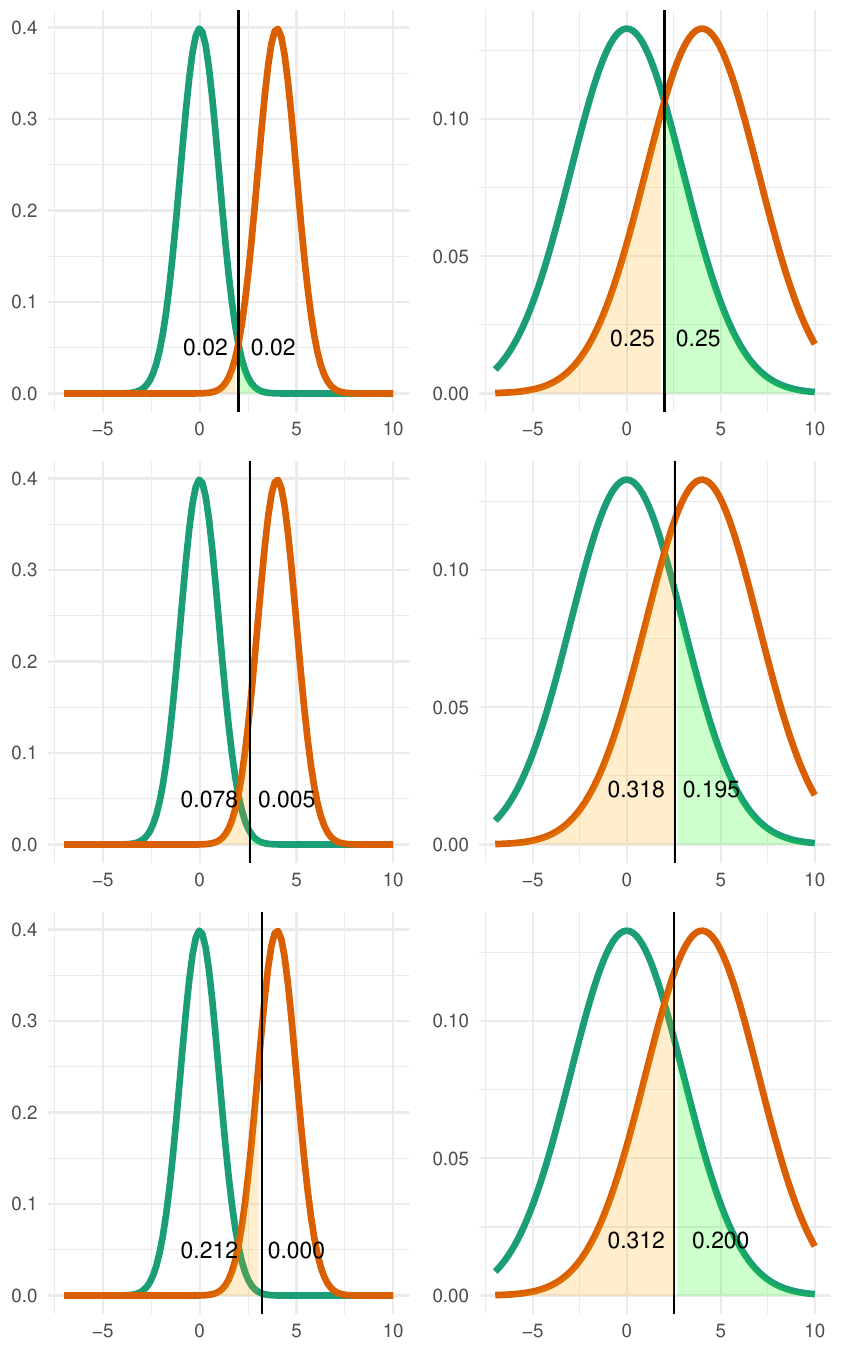}
    \caption{\footnotesize{Plots of three classifiers in Example \ref{ex:simple}. The three rows, from top to bottom, represent figure illustration of the Bayes classifier, NP oracle classifier, and NP-EO oracle classifier, respectively. The left panel illustrates the densities of $X^{0,a}$ and $X^{1,a}$ and the right panel those of $X^{0,b}$ and $X^{1,b}$. In every sub-figure, the green curve represents class $0$ density, and the orange curve represents class $1$ density. In each row, the two thresholds of the classifier are indicated by the two black vertical lines. The type I and type II errors conditional on the sensitive attribute are depicted respectively as the light green and light orange regions in every sub-figure with their values marked.}}\label{fig:npeo_example}
\end{center}
\end{figure}
\end{example}


An NP-EO oracle classifier has a nice property: it is invariant to the changes in the proportions of class labels. 
This insight is concretized by the following proposition.

\begin{prop}\label{prop:invariance}
Under conditions of Theorem \ref{thm:oracle}, an NP-EO oracle classifier is invariant to the change in $\p(Y = 0)$ (or equivalently $\p(Y = 1)$), when the distributions of $X \mid (Y =y, S = s)$ (i.e., $X^{y,s}$) and $S \mid (Y = y)$ stay the same for each $y \in \{0, 1\}$ and $s \in \{a, b\}$. 

\end{prop}


\section{Methodology}\label{sec:method}


In this section, we propose two sample-based NP-EO umbrella algorithms. Theorem \ref{thm:oracle} indicates that the density ratios are the best scores, with proper threshold choices. Hence, plugging the density ratio estimates in equation \eqref{eq:oracle_classifier} would lead to classifiers with good theoretical properties. In practice and more generally, however, practitioners can and might prefer to use scores from canonical classification methods (e.g., logistic regression and neural networks to avoid estimating high-dimensional densities), which we also refer to as \emph{base algorithms}. 
Inspired by  \eqref{eq:oracle_classifier}, we construct classifiers
\begin{align}\label{eq:classifier}
    \hat{\phi}(X, S) = \1\{T^a(X) > c_a\}\cdot \1\{S = a\} + \1\{T^b(X) > c_b\}\cdot \1\{S = b\}\,,
\end{align}
where $T^a(\cdot)$ and $T^b(\cdot)$ are given scoring functions for groups $S=a$ and $S=b$, respectively, and our task is to choose proper data-driven thresholds $c_a$ and $c_b$ that take into account the NP and EO constraints.  This form is inspired by the NP-EO oracle classifier in the previous section by the density ratios with $T^a$ and $T^b$.  
We leave the more theory-oriented investigation on density ratio plug-ins for the future.

The classifier $\hat{\phi}$ in \eqref{eq:classifier} is trained on a finite sample; thus, it is random due to the randomness of the sample, and the constraints in program  \eqref{eq:NP_EO} cannot be satisfied with probability $1$ in general. Therefore, we aim to achieve high-probability NP and EO constraints as follows,
\begin{align}\label{cond:hp-NPEO}
    \p\left(R_0(\hat{\phi}) > \alpha\right) \leq \delta \quad \text{and} \quad \p\left(L_1(\hat{\phi}) > \varepsilon\right) \leq \gamma\,,
\end{align}
for pre-specified small $\delta, \gamma\in(0,1)$. Here, $\p$ is taken over the training sample. 

In Sections \ref{sec:npeo_algorithm} and \ref{sec:npeo_cr}, we will present two umbrella algorithms: $\text{NP-EO}_\text{OP}$ and $\text{NP-EO}_\text{MP}$. The meaning of their names will become clear later. $\text{NP-EO}_\text{OP}$ is simpler and computationally lighter than $\text{NP-EO}_\text{MP}$. It is also ``safer'' in the sense that it achieves at least $1-\delta$ probability type I error control whereas $\text{NP-EO}_\text{MP}$ is only theoretically guaranteed to achieve at least $1 - \delta^+$ probability control for some $\delta^+ \searrow \delta$ as the sample size grows. However, $\text{NP-EO}_\text{OP}$ sacrifices the power. In contrast, $\text{NP-EO}_\text{MP}$ achieves smaller type II error and does not violate exact high-probability NP constraint in numerical analysis, as demonstrated in Section \ref{sec:num}. Moreover, $\text{NP-EO}_\text{MP}$ is a generalization of $\text{NP-EO}_\text{OP}$ in terms of threshold selection. Thus, it is convenient for readers to encounter $\text{NP-EO}_\text{OP}$ first.

\subsection{The $\text{NP-EO}_\text{OP}$ umbrella algorithm}\label{sec:npeo_algorithm}

We now construct an algorithm that respects \eqref{cond:hp-NPEO}\textsuperscript{\footnotemark{}}\footnotetext{\ Strictly speaking, we only achieve $\gamma^+$ in \eqref{cond:hp-NPEO}, where $\gamma^+\searrow \gamma$ as sample size diverges.}, and achieves type II error as small as possible. Denote by $\mathcal{S}^{y,s}$ the set of $X$ feature observations whose labels are $y$ and sensitive attributes are $s$, where $y \in \{0,1\}$ and $s \in \{a,b\}$. We assume that all the $\mathcal{S}^{y,s}$'s are independent, and instances within each $\mathcal{S}^{y,s}$ are i.i.d.  Each $\mathcal{S}^{y,s}$ is divided into two halves: $\mathcal{S}^{y,s}_{\text{train}}$ for training scoring functions, and $\mathcal{S}^{y,s}_{\text{left-out}}$ for threshold estimation in \eqref{eq:classifier}. 

First, all $\mathcal{S}^{y,s}_{\text{train}}$'s are combined to train a scoring function (e.g., sigmoid function in logistic regression) $T: \mathcal{X} \times \{a, b\} \mapsto \R$; then we take $T^a(\cdot) = T(\cdot, a)$ and $T^b(\cdot) = T(\cdot, b)$. To determine $c_a$ and $c_b$, we select pivots to fulfill the NP constraint first and then adjust them for the EO constraint. A prior result leveraged to achieve the high-probability NP constraint is the \emph{NP umbrella algorithm} developed by \cite{tong2018neyman}. This algorithm adapts to all scoring-type classification methods (e.g., logistic regression and neural nets), which we now describe. 
For an arbitrary scoring function $S: \mathcal{X} \mapsto \R$ and i.i.d. class $0$ observations $\{X_1^0, X_2^0, \cdots, X_n^0\}$,  a classifier that controls type I error under $\alpha$ with probability at least $1 - \delta$ and achieves small type II error can be built as
$
    \1\{S(X) > s_{(k^*)}\},
$
where $s_{(k^*)}$ is the $(k^*)^{\text{th}}$ order statistic of $\{s_1, s_2, \cdots, s_n\} := \{S(X_1^0), S(X_2^0), \cdots, S(X_n^0)\}$ and $k^*$ is the smallest $k\in\{1,2,\cdots,n\}$ such that
$
    \sum_{j=k}^n{n \choose j}(1-\alpha)^j\alpha^{n-j} \leq \delta.
$ The smallest such $k$ is chosen for type II error minimization. The only condition for this high-probability type I error control is $n \geq \lceil\log\delta/\log(1 - \alpha)\rceil$, a mild sample size requirement. Details of this algorithm are recollected from \cite{tong2018neyman} and provided in Supplementary Materials B.1. 

Motivated by the NP umbrella algorithm, we apply $T^s(\cdot)$ to each instance in $\mathcal{S}^{y,s}_{\text{left-out}}$ to obtain $\mathcal{T}^{y,s} = \{t^{y,s}_1, t^{y,s}_2, \cdots, t^{y,s}_{n^y_s}\}$, where $n^y_s = |\mathcal{S}^{y,s}_{\text{left-out}}|$, $y\in\{0,1\}$, and $s\in\{a,b\}$. A natural starting point is to apply the NP umbrella algorithm \citep{tong2018neyman} to the data with sensitive attributes $a$ and $b$ separately so that they both satisfy the NP constraint \eqref{cond:hp-NPEO}.  Concretely, from the sorted set $\mathcal{T}^{0,a} = \{t^{0,a}_{(1)}, t^{0,a}_{(2)},\cdots,t^{0,a}_{(n^0_a)}\}$, the pivot $t^{0,a}_{(k^{0,a}_*)}$ is selected as the $\left(k^{0,a}_*\right)^{\text{th}}$ order statistic in $\mathcal{T}^{0,a}$, where $k^{0,a}_*$ is the smallest $k\in\{1, \cdots, n^0_a\}$ such that
$
    \sum_{j = k}^{n^0_a}{n^0_a \choose j}(1 - \alpha)^j\alpha^{n_a^0-j} \leq \delta.
$
The pivot $t^{0,b}_{(k^{0,b}_*)}$ is selected similarly on $\mathcal{T}^{0,b}$. If $c_a \geq t^{0,a}_{(k^{0,a}_*)}$ and $c_b \geq t^{0,b}_{(k^{0,b}_*)}$, then the classifier $\hat{\phi}$ in \eqref{eq:classifier} satisfies 
\begin{align} \label{eq:fan1}
    \p\left(R_0^a(\hat{\phi}) > \alpha\right) \leq \delta \quad \text{ and }\quad  \p\left(R_0^b(\hat{\phi}) > \alpha\right) \leq \delta\,,
\end{align}
by Proposition 1 in \cite{tong2018neyman}. In view of \eqref{eq: error decomposition}, the above inequalities guarantee that the NP constraint can be achieved with probability at least  $1-2\delta$. 
If we want to strictly enforce the $1 - \delta$ probability type I error control in theory as in inequality \eqref{cond:hp-NPEO}, the $\delta$ parameter in our algorithm can be replaced by $\delta/2$\textsuperscript{\footnotemark{}}\footnotetext{\ However, numerical results in Section \ref{sec:num} suggest that this extra cautionary measure does not seem to be necessary in practice, because the subsequent EO adjustment step gears our algorithm towards the more conservative direction for type I error control.}. 

The next step is to adjust the thresholds so that the resulting classifier also satisfies EO inequality in \eqref{cond:hp-NPEO}, i.e., the high-probability EO constraint. To keep the NP constraint, we increase the values of thresholds for both groups.  
Similar to $\mathcal{T}^{0,a}$ and $\mathcal{T}^{0,b}$, we denote the sorted $\mathcal{T}^{1,s} = \{t^{1,s}_{(1)}, t^{1,s}_{(2)},\cdots,t^{1,s}_{(n^1_{s})}\}$ for $s \in \{a, b\}$ and select $c_a$ from $\mathcal{T}^{1,a}$ and $c_b$ from $\mathcal{T}^{1,b}$ in order to facilitate the type II error calculation. 
Let
\begin{align}\label{eq:l_a}
    l_a = \sum_{j=1}^{n^1_a} \1\left\{t^{1,a}_j \leq t^{0,a}_{(k^{0,a}_*)}\right\} \quad \text{ and }\quad   l_b = \sum_{j=1}^{n^1_b} \1\left\{t^{1,b}_j \leq t^{0,b}_{(k^{0,b}_*)}\right\} \,.
\end{align}
Then, $c_a$ is selected from $\{t^{1,a}_{(j)}: l_a< j \leq n^1_a\}$ and $c_b$ is selected from $\{t^{1,b}_{(j)}: l_b < j \leq n^1_b\}$ so that 
\eqref{eq:fan1} holds.   To this end, we investigate the distributions of
\begin{align*}
    r^a_1(i) = \p_{X^{1,a}}\left(T^a(X^{1,a}) \leq t^{1,a}_{(i)} \right)\quad \text{ and }\quad 
r^b_1(j) = \p_{X^{1,b}}\left(T^b(X^{1,b}) \leq t^{1,b}_{(j)}\right)\,,
\end{align*}
for $i > l_a$ and $j > l_b$.  They are respectively the $S = a$ and $S = b$ components of the type II error of the classifier in \eqref{eq:classifier}, if we take $c_a = t^{1,a}_{(i)}$ and $c_b = t^{1,b}_{(j)}$; they are random because only the randomness of $X^{1,a}$ and $X^{1,b}$ are taken in $\p_{X^{1,a}}$ and $\p_{X^{1,b}}$. We need to understand these two quantities to choose from all eligible pairs $i$ and $j$ that satisfy the EO constraint.

The left-hand side of the EO inequality in \eqref{cond:hp-NPEO} can be written as
$
    \p\left(\left|r^a_1(i) - r^b_1(j)\right| > \varepsilon\right),
$
since we can consider the scoring function $T (\cdot)$ (and hence $T^a(\cdot)$ and $T^b(\cdot)$) as fixed due to independent pre-training of $T(\cdot)$.   Since the random variables $r^a_1(i)$ and $r^b_1(j)$ are independent and admit similar definitions, we need only to study one of them as follows. 

Let $X$ and $Y_1, Y_2, \cdots, Y_n$ be continuous, independent, and identically distributed random variables. Moreover, let $c$ be a random variable that is independent of $X, Y_1, \cdots, Y_n$ and define by $l = \sum_{j=1}^n \1\{Y_j \leq c\}$.   Our goal is to approximate the distribution of $\p_X(X \leq Y_{(k)})$ conditional on $l$ for $k > l$, which is needed for $r^a_1(i)$  and $r^b_1(j)$.   Note that the conditional probability does not depend on the original distribution of $X$ and
$$
 \p_X(X \leq Y_{(k)} \mid l) = \p_X(X \leq Y_{(l)} \mid l) +  \p_X( Y_{(l)} <  X \leq Y_{(k)} \mid l)\,.
$$

By using the property of the uniform order statistics, it can be shown that the above quantity
 has the same distribution as $g_{c,l} + \l(1 - g_{c,l}\r)B_{k-l, n-k+1}$ for $k > l$ 
with independent random variables $g_{c,l} = \p(Y_1 \leq c \mid l)$ and $B_{k-l,n-k+1} \sim \text{Beta}(k-l,n-k+1)$.  It remains to approximate the distribution of $g_{c,l}$, which is $l/n$ if $c$ is a constant.  Recall that $c$ is a random variable and $g_{c, l}= \mathbb E(F(c)|l)$ where $F$ is the cdf of $Y_1$.  Writing $\theta = F(c)$,  from the Bayesian point of view, the distribution of $g_{c,l}$ is the posterior distribution of $\theta$ given $n$ i.i.d. Bernoulli($\theta$) observations with sufficient statistic $l$. By Bernstein-von Mises theorem, $g_{c,l}$ is ``close'' to being normally distributed with mean $l/n$ (MLE in frequentist view) and variance equal to the Fisher information of the Bernoulli trial at MLE: $n^{-1}(l/n)(1-l/n)$.

\begin{SCfigure}[][h]
     \centering
     \includegraphics[width = 0.6\textwidth, height = 6cm]{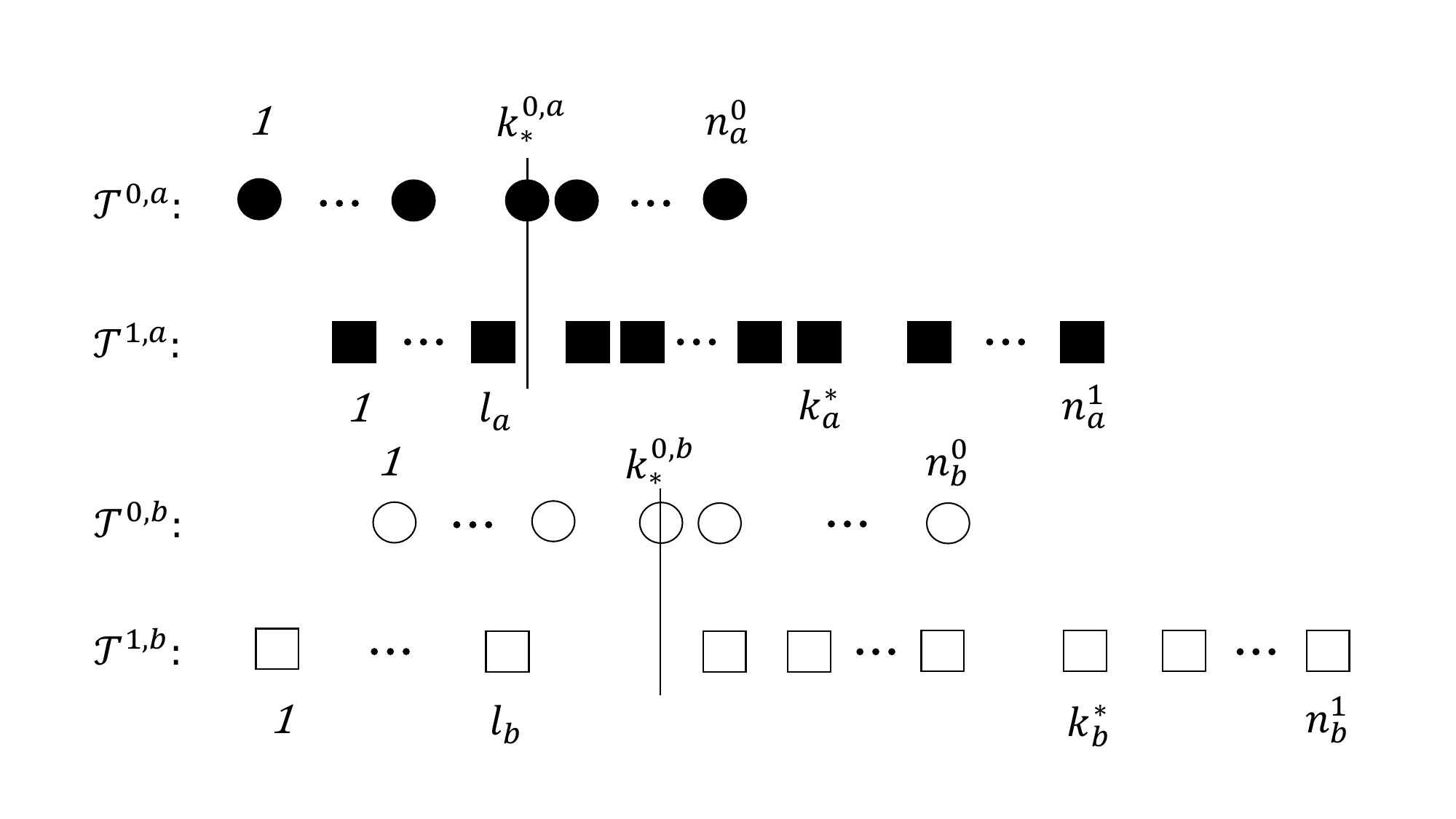}
     \vspace{-2.2em}
    \caption{\footnotesize{A cartoon illustration of the choices of $k_a^*$ and $k_b^*$.  They are moved in the NP-constrained feasible region (to the left) to search for the pairs that satisfy the EO constraint and to pick the most powerful pair. For every $\mc T^{y,s}$, the circles, or squares, in their corresponding row, represent their sorted elements, ascending from left to right. \label{cartoon illustration}}}
   \end{SCfigure}

The above discussion reveals that  the distribution of  $(r^a_1(i) \mid l_a)$
can be approximated by $G^{1,a} + (1 - G^{1,a})B_{i-l_a, n^1_a - i + 1}$ where
$
    G^{1,a} \sim \mathcal{N}\l(\frac{l_a}{n^1_a}, \frac{l_a/n^1_a(1 - l_a/n^1_a)}{n^1_a}\r).
$
Similarly, the distribution of $(r_1^b(j) \mid l_b)$ can be approximated. 
Let $F^{1,a}(i)$ and $F^{1,b}(j)$ be two independent random variables such that
$
    F^{1,a}(i) = G^{1,a} + (1 - G^{1,a})B_{i-l_a, n^1_a - i + 1},
$
in distribution and $F^{1,b}(j)$ is defined analogously. Then, we can pick $(i,j)$ such that
\begin{align}\label{eq:selection_criterion}
    \p\l(\l|F^{1,a}(i) - F^{1,b}(j)\r| > \varepsilon\r) \leq \gamma\,.
\end{align}
Among these feasible pairs, the one that minimizes the empirical type II error, which can be calculated as $\l((i - 1) + (j - 1)\r) / (n_a^1 + n_b^1)$, should be selected; i.e., we select 
\begin{align}\label{eq:min_empirical_type_II}
    (k^*_a,k^*_b) = \min_{\text{all feasible } (i,j)  \text{ that satisfy} \eqref{eq:selection_criterion}} \frac{i  + j- 2}{n_a^1 + n_b^1}\,.
\end{align}
The process to arrive at $(k^*_a,k^*_b)$ is illustrated in Figure \ref{cartoon illustration}.  We propose an NP-EO classifier
\begin{align*}
    \hat{\phi}^*(X,S) = \1\{T^a(X) > t^{1,a}_{(k^*_{a})}\}\cdot\1\{S = a\} + \1\{T^b(X) > t^{1,b}_{(k^*_{b})}\}\cdot\1\{S = b\}\,.
\end{align*}
Note that, if none of  $i\in\{l_a+1, \cdots, n^1_a\}$ and $j\in\{l_b+1, \cdots, n^1_b\}$ satisfy inequality \eqref{eq:selection_criterion}, we say our algorithm does not provide a viable NP-EO classifier.  This kind of exception has not occurred in simulation or real data studies. 

We summarize the above NP-EO umbrella algorithm in Algorithm \ref{alg:main_alg}. Note that in Step 8, the NP violation rate control at $\delta / 2$ is needed for theoretical purposes (c.f. Theorem \ref{thm:NP-EO-classifier} and its proof). We will demonstrate through numerical analysis that it suffices to use $\delta$ instead. 
We also note that the steps to reach $(k^*_a, k^*_b)$ are summarized as the  \emph{EO violation algorithm} (Step 10) inside Algorithm \ref{alg:main_alg}, also presented separately as Algorithm 4 in Supplementary Materials for clarity. The next theorem provides a theoretical guarantee for $\hat{\phi}^*(X,S)$. 

\begin{algorithm}[h!!!]
\caption{$\text{NP-EO}_\text{OP}$ umbrella algorithm [``OP" stands for One (pair of) Pivots] \label{alg:main_alg}}
\SetKw{KwBy}{by}
\SetKwInOut{Input}{Input}\SetKwInOut{Output}{Output}
\SetAlgoLined

\Input{\footnotesize $\mathcal{S}^{y,s}$: $X$ observations whose label $y\in\{0, 1\}$ and sensitive attribute $s\in\{a, b\}$ \\
$\alpha$: upper bound for type I error\\
$\delta$: type I error violation rate target \\
$\varepsilon$: upper bound for the type II error disparity \\
$\gamma$: type II error disparity violation rate target
}

\footnotesize $\mathcal{S}^{y,s}_{\text{train}}, \mathcal{S}^{y,s}_{\text{left-out}} \leftarrow$ random split on $\mathcal{S}^{y,s}$ for $y \in \{0, 1\}$ and $s \in \{a,b\}$

\footnotesize $\mathcal{S}_{\text{train}} \leftarrow \mathcal{S}^{0,a}_{\text{train}} \cup \mathcal{S}^{0,b}_{\text{train}} \cup \mathcal{S}^{1,a}_{\text{train}} \cup
\mathcal{S}^{1,b}_{\text{train}}$

\footnotesize $T \leftarrow$ \textsf{base classification algorithm}$(\mathcal{S}_{\text{train}})$ \tcp*{$T(\cdot,\cdot): \mathcal{X} \times \{a,b\} \mapsto \R$} 

\footnotesize $T^s(\cdot) \leftarrow T(\cdot, s)$ for $s \in \{a,b\}$

\footnotesize $\mathcal{T}^{y,s} \leftarrow T^s(\mathcal{S}^{y,s}_{\text{left-out}})$ for $y \in \{0,1\}$ and $s \in \{a,b\}$

\footnotesize $n^y_s \leftarrow |\mathcal{T}^{y,s}|$ for $y \in \{0,1\}$ and $s \in \{a,b\}$

\footnotesize $\mathcal{T}^{y,s} = \{t^{y,s}_{(1)}, t^{y,s}_{(2)},\cdots,t^{y,s}_{(n^y_{s})}\}$ for $y \in \{0,1\}$ and $s \in \{a,b\}$

\footnotesize $k^{0,s}_* \leftarrow \textsf{the NP umbrella algorithm}(n^{0}_s, \alpha, \delta/2)$ for $s \in \{a,b\}$ 

\footnotesize $l_s \leftarrow \max\{k \in \{1,2,\cdots,n_{s}^1\}: t^{1,s}_{(k)} \leq t^{0,s}_{(k^{0,s}_*)}\}$ for $s \in \{a,b\}$

\footnotesize $k^*_a, k^*_b \leftarrow \textsf{EO violation algorithm}(l_a, l_b, n_a^1, n_b^1, \varepsilon, \gamma)$ in Supplementary Materials G.

\Output{\footnotesize $\hat{\phi}^*(X,S) = \1\{T^a(X) > t^{1,a}_{(k^*_{a})}\}\cdot\1\{S = a\} + \1\{T^b(X) > t^{1,b}_{(k^*_{b})}\}\cdot\1\{S = b\}$}
\end{algorithm}

\begin{theorem}\label{thm:NP-EO-classifier}
Let $\hat{\phi}^*(\cdot, \cdot)$ be the classifier output by Algorithm \ref{alg:main_alg} with parameters $(\alpha, \delta/2, \varepsilon, \gamma)$. Assume that the scoring function $T(\cdot, \cdot)$ is trained such that $T^s(X^{y,s})$ is a continuous random variable whose distribution function is strictly monotone for each $y \in\{0,1\}$ and $s \in\{a,b\}$, and that all distribution functions for $T^s(X^{y,s})$ have the same support. Furthermore, assume that $n^0_a,n^0_b, n^1_a, n^1_b$ are deterministic and $\min\{n^0_a,n^0_b\} \geq \log (\delta/2) / \log (1 - \alpha)$. 
Then, it holds simultaneously that
\begin{align*}
      \text{(a) }   \p\l(R_0(\hat{\phi}^*) > \alpha\r) \leq \delta  \quad \text{ and }\quad  \text{(b) } \p\left(|R_1^a(\hat{\phi}^*) - R_1^b(\hat{\phi}^*)| > \varepsilon\right) \leq \gamma + \xi(n_a^1, n_b^1)\,,
\end{align*}
in which $\xi(n^1_a, n^1_b)$ converges to $0$ as $n_a^1$ and $n_b^1$ diverge.
\end{theorem}

 In Theorem \ref{thm:NP-EO-classifier}, the conditions for distributions of $T^s(X^{y,s})$ ensure that the Bernstein-von Mises theorem can be invoked and is often met in practice. For example, when all features are continuous, and the base model is a probabilistic algorithm such as logistic regression, the resulting scores satisfy the continuity assumption and have a range of $(0,1)$ for all values of $s$ and $y$. The second assumption is the typical sample size requirement as in \cite{tong2018neyman}, ensuring that the sample size is sufficient for the NP umbrella algorithm to select a threshold that meets the high-probability NP constraint. For instance, when $\alpha=\delta=0.1$, it requires $\min\{n^0_a,n^0_b\} \geq \log (\delta/2) / \log (1 - \alpha)=28.43$, which is not a very stringent in practice.

Indeed, take the $S = a$ component for example, this theorem is applied to the class of binomial sample $l_a$ defined in \eqref{eq:l_a}, whose probability of success is $\p_{X^{1,a}}\left(T^a(X^{1,a}) \leq t^{1,a}_{(i)} \right)$. The key issue here is that this random probability needs to be in the interior of $[0, 1]$ with probability $1$, which is guaranteed by assumptions on $T^s(X^{y,s})$. Next, the assumptions for $n^0_a$ and $n^0_b$, adapted from \cite{tong2018neyman}, are mild sample size requirements to ensure the high-probability NP constraint (c.f. part (a) of Theorem \ref{thm:NP-EO-classifier}). 
We note that part (b) of Theorem \ref{thm:NP-EO-classifier} states that the type II error disparity violation rate can be controlled by $\gamma$ plus a term that vanishes asymptotically. This extra term, asymptotically negligible, is the price for the errors of Gaussian approximation on the distributions of $r_1^a$ and $r_1^b$. 

\subsection{The $\text{NP-EO}_\text{MP}$ umbrella algorithm}\label{sec:npeo_cr}

{\color{black}
In this section, we briefly introduce a variant of Algorithm \ref{alg:main_alg}. For the rest of the manuscript, Algorithm \ref{alg:main_alg} will be referred to as $\text{NP-EO}_\text{OP}$, where ``OP'' stands for \textbf{O}ne (pair of) \textbf{P}ivots, and the variant, which relies on \textbf{M}ultiple (pairs of) \textbf{P}ivots, will be referred to as $\text{NP-EO}_\text{MP}$.

By selecting one pair of pivots as lower bounds for threshold candidates, the $\text{NP-EO}_\text{OP}$ algorithm follows a ``conservative'' approach;  it ensures that thresholds $c_a$ and $c_b$ are chosen such that $R^a_0$ and $R^b_0$ are both controlled by $\alpha$ with high probability, whereas we only need $R_0$, a weighted average of $R^a_0$ and $R^b_0$, to be controlled under $\alpha$. Hence, the sensitive-attribute conditional type I error control in $\text{NP-EO}_\text{OP}$ is not necessary to meet NP constraint and may lead to unnecessarily small $R_0$ (and large $R_1$). If we can relax the control on $R^a_0$ and $R^b_0$ while still maintaining control on $R_0$, we could enhance the classifier's power.


Following this idea, $\text{NP-EO}_\text{MP}$ chooses multiple pairs of pivots to cater to the NP constraint, with each pair serving as lower bounds for threshold candidates.  These pairs include the one in the $\text{NP-EO}_\text{OP}$ algorithm, rendering $\text{NP-EO}_\text{OP}$ effectively a special case of $\text{NP-EO}_\text{MP}$. Each pivot pair generates multiple pairs of thresholds, leading to a set of potential classifiers. Then, we choose among all the potential classifiers the one that (approximately) minimizes the empirical type II error. 

This approach faces three challenges compared to the $\text{NP-EO}_\text{OP}$ algorithm: (a) in the $\text{NP-EO}_\text{MP}$ algorithm, without separate control over $R_0^a$ and $R_0^b$, it is necessary to identify all pivot pairs that can control $R_0$ with high probability; (b) the estimation of type II error disparity by the $\text{NP-EO}_\text{OP}$ algorithm depends on approximating the distributions of $R_0^a$ and $R_0^b$ for all potential classifiers. These distributions are characterized by parameters that involve $l_a$ and $l_b$ as defined in \eqref{eq:l_a}, representing the ranks of the $S=a$ and $S=b$ pivots among the class $1$ scores for $S=a$ and $S=b$. On the other hand, the multiple pivot pairs and thereby ranks in the $\text{NP-EO}_\text{MP}$ do not align with the original distributional setting. Consequently, the method employed in the $\text{NP-EO}_\text{OP}$ algorithm becomes invalid; (c) the $\text{NP-EO}_\text{MP}$ algorithm generates a significantly larger number of potential classifiers compared to the $\text{NP-EO}_\text{OP}$ algorithm. Selecting the classifier that minimizes empirical type II error can be computationally inefficient. 

To address the first challenge, note that the pivot pair determines the upper bound for empirical type I errors of its implied potential classifiers by the proportions of class $0$ observations exceeding their respective pivot values. Thus, in the $\text{NP-EO}_\text{MP}$ algorithm, we only investigate the pair of pivots that can achieve the same empirical type I error upper bound as the pivot pair in the $\text{NP-EO}_\text{OP}$ algorithm. Essentially, when the sample size is large, the discrepancy between empirical and population-level type I errors should be uniformly small across all potential classifiers. Therefore, matching upper bounds for empirical type I errors in both $\text{NP-EO}_\text{OP}$ and $\text{NP-EO}_\text{MP}$ imply similar upper bounds for population-level type I errors, and thus simultaneous high probability type I error control. To tackle the second issue, we employ an extended Gaussian approximation of posteriors involving all pivots to approximate the distributions of $R_0^a$ and $R_0^b$ of all potential classifiers. For the third challenge, rather than examining all potential classifiers, we develop an adaptive approach to reduce search time. The exact algorithm construction is quite involved. For a complete description, please refer to Algorithm 2 in Section A in Supplementary Materials. Next, we present the theoretical guarantee for $\text{NP-EO}_\text{MP}$.
}

\begin{theorem}\label{thm:adj-NP-EO-classifier}
    Let $\hat{\phi}^{**}(\cdot, \cdot)$ be the classifier output by Algorithm 2 in Supplementary Materials with parameters $(\alpha - \eta, \delta, \varepsilon, \gamma)$ for $0<\eta \ll \alpha$. Assume that the scoring function $T(\cdot, \cdot)$ is trained such that the same conditions in Theorem \ref{thm:NP-EO-classifier} hold. Moreover, for each $y \in \{0,1\}$, define $n^y = n^y_a + n^y_b$ and suppose $n^y_a$ is independently binomial distributed with size $n^y$ and success rate $p_{a \mid y}$. Assume  that $n^0 \geq \log(\delta) / \log(1 - \alpha)$. Then, it holds simultaneously that 
        \begin{align*}
            &(a)\quad \p\l(R_0(\hat{\phi}^{**}) > \alpha\r) \leq \delta +  2e^{-\frac{1}{32}n^0(p_{a|0} - \eta/8)\eta^2} + 2e^{-\frac{1}{32}n^0(p_{b|0} - \eta/8)\eta^2} + 2e^{-\frac{1}{32}n^0\eta^2} + 2e^{-\frac{1}{2}n^0\eta^2}\,,\\
       &(b)\quad \p\left(|R_1^a(\hat{\phi}^{**}) - R_1^b(\hat{\phi}^{**})| > \varepsilon\right) \leq \gamma + \xi'(n^1)\,,
        \end{align*}
in which $\xi'(n^1)$ converges to $0$ as $n^1$ diverges.
\end{theorem}

The proof of this theorem is presented in the Supplementary Materials. Here, we remark that the main difference between Theorems \ref{thm:NP-EO-classifier} and \ref{thm:adj-NP-EO-classifier} is in part (a). In Theorem \ref{thm:NP-EO-classifier}, the type I error is controlled with probability at least $1 - \delta$, whereas in Theorem \ref{thm:adj-NP-EO-classifier}, $\hat{\phi}^{**}$ only gives an ``approximately'' $1 - \delta$ type I error control. This is not surprising since {\color{black} we relax the strict approach of separately controlling $R_0^a$ and $R_0^b$ and thus can only achieve an ``approximate'' control of $R_0$. This yields the exponential terms in part (a) of Theorem \ref{thm:adj-NP-EO-classifier}.}

\section{Numerical results}\label{sec:num}

In this section, we present simulation and real-data evidence that supports the effectiveness of the newly proposed NP-EO algorithms. 
In each simulation setting, all trained algorithms are evaluated on a large test set to approximate the (population-level) type I and type II errors. This procedure is repeated $1{,}000$ times, and thus $1{,}000$ copies of (approximate) type I and type II errors can be acquired. Then, the NP violation rate is computed as the proportion of type I errors exceeding the target level defined in the NP constraint. Similarly, the EO violation rate is computed as the proportion of type II error disparity exceeding the target level defined in the EO constraint. Finally, recall that for $\text{NP-EO}_{\text{OP}}$ algorithm, we use $\delta$, instead of $\delta/2$, in Algorithm \eqref{alg:main_alg}.

In particular, Simulation 1 examines the comparison between $\text{NP-EO}_{\text{OP}}$, $\text{NP-EO}_{\text{MP}}$ and two established classifiers that incorporate fairness constraints: FairBayes \citep{zeng2022bayes} and CSL, which is developed within the cost-sensitive learning framework \citep{menon2018cost}. Additionally, we highlight the trade-off between fairness and efficiency by observing the relationship between type II error and EO constraints when we vary $\alpha$ and $\varepsilon$, which is further compared with CSL. Simulation S.2 in the Supplementary Material F.3 focuses on the comparison between $\text{NP-EO}_{\text{OP}}$, $\text{NP-EO}_{\text{MP}}$ and other NP classifiers. 

\subsection{Simulation}\label{sec:simulation}

In all settings, for each $y\in\{0,1\}$ and $s\in\{a,b\}$, we generate $n^{y,s}$ training observations and $100n^{y,s}$ test observations. We evaluate the performance of the $\text{NP-EO}_\text{OP}$ and $\text{NP-EO}_\text{MP}$ against two existing methods: FairBayes and CSL in Simulation \ref{sim:normal}, and the classical algorithm, the NP umbrella algorithm, and the NP umbrella algorithm mixed with random guesses in Simulation S.2 in Supplementary Material F.3.

Notably, FairBayes and CSL are not inherently designed under the NP-EO paradigm, necessitating distinct parameter configurations. Specifically, FairBayes requires only the EO constraint control level $\varepsilon$, without the need for NP-related parameters ($\alpha$ and $\delta$) or the EO violation rate target $\gamma$. On the other hand, CSL \citep{menon2018cost} is more subtle. It  assigns distinct costs $c$, $\Bar{c}$, and $\lambda$ to $R_0$, $R_1$, and $L_1$, respectively. However, these costs cannot be directly mapped to $\alpha$ and $\varepsilon$ within the NP-EO framework. To achieve a fair comparison,  we exhaustively explore combinations of $(c, \Bar{c}, \lambda)$ through binary search based on a training dataset of size $20,000$. We then select the cost combination that minimizes $\widehat{R}_1$ while ensuring $\widehat{R}_0 \leq \alpha$ and $\widehat{L}_1 \leq \varepsilon$\footnote{Here, we utilize a large test data of size 20,000 to approximate the population.}. The chosen combination is then applied in subsequent analyses in an independently generated dataset.  The classical algorithm (e.g. logistic regression, support vector machines) serves as the baseline algorithm without any adjustments for NP or EO constraints. The NP umbrella algorithm adjusts the base algorithms for the NP constraint and is described in Section B.1. A detailed description of the NP umbrella algorithm mixed with random guess can be found in Supplementary Material F.3.

\begin{simulation}\label{sim:normal}
Let $X^{y,s}$ be multidimensional Gaussian with mean $\mu_{y,s}$ and covariance matrix $\Sigma_{y,s}$ for each $y \in \{0,1\}$ and $s \in \{a,b\}$. Here, $\mu_{0,a} = (0,1,1)^\top$,  $\mu_{1,a} = (0,0,0)^\top$, $\mu_{0,b} = (0,0,3)^\top$ and $\mu_{1,b} = (1,0,-1)^\top$. Moreover, $\Sigma_{y,s}$ is $2I$ where $I$ is the identity matrix for every $y \in \{0, 1\}$ and $s \in \{a,b\}$. Furthermore, $n^{0,a} = 800$, $n^{1,a} = 1200$, $n^{0,b} = 800$ and $n^{1,b} = 1200$. In this setting, $\alpha$ and $\varepsilon$ both have varying values. We also set $\delta = 0.05$ and $\gamma = 0.05$. The base algorithm used is logistic regression. The results are reported in Table \ref{tb::simulation 1} and Figure 1 - 3 in the Supplementary Material.  

\end{simulation}

As shown in Table \ref{tb::simulation 1}, across various combinations of $\alpha$ and $\varepsilon$, both $\text{NP-EO}_{\text{OP}}$ and $\text{NP-EO}_{\text{MP}}$ effectively satisfy the NP and EO constraints with the targeted high probability. In contrast, the FairBayes algorithm does not incorporate NP control ($\alpha$) or the high probability bounds defined by $\delta$ and $\gamma$. As a result, FairBayes fails to maintain type I error control at the desired level. More specifically, it exhibits an NP violation rate close to 1 across all settings.

On the other hand, with CSL, we observe that the average $R_0$ and average $L_1$ are effectively controlled at $\alpha$ and $\varepsilon$, respectively, due to our strategic choice of costs. However, the observed NP and EO violation rates range from $0.333$ to $0.883$, substantially exceeding the target thresholds of $\delta=\gamma=0.05$. This outcome aligns with expectations, as the cost-sensitive learning framework inherently lacks mechanisms for high probability guarantee - its primary focus is empirical risk minimization. In summary, while FairBayes and CSL are established methods for achieving fairness, $\text{NP-EO}_{\text{OP}}$ and $\text{NP-EO}_{\text{MP}}$ offer superior performance when the goal is to satisfy both NP and EO constraints with specific high probabilities.

Lastly, Figure 3 in the Supplementary Material offers a clear view of the interplay between $\varepsilon$ and $\alpha$. Examining the rows, we observe that at each fixed $\alpha$, as we increase $\varepsilon$ (relaxing the EO constraint target) from $0.1$ to $0.2$, the average $R_1$ decreases and the average $L_1$ increases, as expected. For each fixed $\varepsilon$, as $\alpha$ (the NP constraint) increases, the average $R_1$ consistently decreases.  Figure 1 in the Supplementary material plots $R_1$ and $L_1$ directly as the two axes to better illustrate the trade-off between them as $\varepsilon$ varies. While $\text{NP-EO}_{\text{OP}}$ and $\text{NP-EO}_{\text{MP}}$ maintain perfect control over the average $L_1$ values for all $\varepsilon$ (the EO constraint), CSL exhibits slightly elevated $L_1$ averages compared to the corresponding targets $\varepsilon$, when $\alpha=0.2$. Notably, CSL achieves smaller average $R_1$ than both $\text{NP-EO}_{\text{OP}}$ and $\text{NP-EO}_{\text{MP}}$, consistent with its lack of control over $R_0$ and $L_1$ constraints with high probability.

\begin{table}[h!]
\centering
\caption{\footnotesize Averages of type I/II errors, along with violation rates of the NP and EO constraints over $1{,}000$ repetitions for Simulation 1. Standard error of the means ($\times 10^{-4}$) in parentheses. For $\text{NP-EO}_{\text{OP}}$ and $\text{NP-EO}_{\text{MP}}$, we set $\delta=\gamma = 0.05$.}\label{tb::simulation 1}
\resizebox{1\linewidth}{!}{
\begin{tabular}{cccccrrrrrrrrrrrrrrrrrrrrrrr}
\hline
 \multicolumn{1}{c}{$\varepsilon$}    & &  algorithms  & &   
 \multicolumn{2}{c}{average $R_0$}   &  & 
 \multicolumn{2}{c}{average $R_1$}   &  &  
 \multicolumn{2}{c}{average $L_1$} &&\multicolumn{2}{c}{NP violation rate}  &&\multicolumn{2}{c}{EO violation rate} \\
 \cline{5-6}  \cline{8-9}   \cline{11-12} \cline{14-15} \cline{17-18}
   && & & $\alpha=.10$ & $\alpha=.20$ & & $\alpha=.10$ & $\alpha=.20$ & & $\alpha=.10$ & $\alpha=.20$ &&$\alpha=.10$ & $\alpha=.20$&&$\alpha=.10$ & $\alpha=.20$\\
\cline{1-1}   \cline{3-3} \cline{5-6}  \cline{8-9}   \cline{11-12} \cline{14-15} \cline{17-18}
\multirow{4}{*}{$\varepsilon=0.1$} & & $\text{NP-EO}_{\text{OP}}$ & & .038(2.1) & .085(3.1) && .648(11.3) & .465(10.8) && .053(8.5) & .052(8.7) && 0(0.0) & 0(0.0)&& .051(69.6) & .047(67.0)\\  
                    & & $\text{NP-EO}_{\text{MP}}$ & & .081(4.0) & .176(4.0) && .492(14.8) & .263(8.6) && .044(9.4) & .045(8.3) && .029(53.1) & .040(62.0)&& .052(70.2) & .023(47.4)\\  
                    & & $\text{FairBayes}$ & & .285(8.2) & .285(8.2) && .139(6.4) & .139(6.4) && .047(5.4) & .047(5.4) && 1(0.0) & .999(10.0) && .007(26.4) &.007(26.4)\\ 
                     & & $\text{CSL}$ & & .102(2.1) & .206(2.1) && .399(6.0) & .193(2.5) && .084(12.2) & .109(4.9) && .639(152.0) & .810(124.1) && .333(149.1) &.724(141.4)\\ 
                    \cline{1-1}   \cline{3-3} \cline{5-6}  \cline{8-9}   \cline{11-12} \cline{14-15} \cline{17-18}
\multirow{3}{*}{$\varepsilon=0.15$} & & $\text{NP-EO}_{\text{OP}}$ & & .038(2.1) & .085(3.2) && .624(11.3) & .440(10.7) && .101(9.0) & .101(9.1) && 0(0.0) & 0(0.0)&& .040(62.0) & .045(65.6)\\ 
                    & & $\text{NP-EO}_{\text{MP}}$ & & .082(3.0) & .177(4.1) && .470(13.5) & .241(8.0) && .073(12.3) & .096(10.0) && .034(57.3) & .048(67.6)&& .022(46.4) & .024(48.4)\\    
                    & & $\text{FairBayes}$ & & .274(9.1) & .274(9.1)&& .143(6.6) & .143(6.6) && .072(8.9) &.072(8.9) && 1(0.0) & .999(10.0) && .002(14.1) &.002(14.1)\\ 
                     & & $\text{CSL}$ & & .103(2.1) & .207(2.1) && .374(5.4) & .181(2.2) && .137(11.1) & .158(4.2) && .675(148.2) & .838(116.6) && .353(151.2) &.747(137.5)\\ 
                   \cline{1-1}   \cline{3-3} \cline{5-6}  \cline{8-9}   \cline{11-12} \cline{14-15} \cline{17-18}
\multirow{3}{*}{$\varepsilon=0.2$} & & $\text{NP-EO}_{\text{OP}}$ & & .038(2.1) & .087(3.2) && .599(11.4) & .414(10.7) && .151(9.1) & .153(9.0) && 0(0.0) & 0(0.0)&& .033(56.5) & .037(59.7)\\  
                    & & $\text{NP-EO}_{\text{MP}}$ & & .082(3.0) & .177(4.2) && .449(14.0) & .223(7.0) && .113(15.0) & .152(8.8) && .032(55.7) & .048(67.6)&& .010(31.5) & .036(58.9)\\  
                    & & $\text{FairBayes}$ & & .271(10.0) & .271(10.0) &&.144(7.1) & .144(7.1) && .079(11.6) & .079(11.6) && 1(0.0) & .990(31.5) && 0(0.0) & 0(0.0)\\ 
                     & & $\text{CSL}$ & & .104(2.1) & .208(2.1) && .352(4.8) & .175(2.0) && .191(9.9) & .205(3.9) && .696(145.5) & .883(101.7) && .380(153.6) &.677(147.9)\\ 
  \hline
			\end{tabular}}
\end{table}

\vspace{-0.5cm}
\subsection{Real data analysis}\label{sec:real_data}
Lenders’ discrimination against a certain social group in the credit market has been a major challenge in financial regulation. Notably, the Equal Credit Opportunity Act in the US, which was enacted in 1974, explicitly makes it unlawful for any creditor to discriminate against applicants based on race, sex, and other non-credit-related social factors. Nevertheless, ample evidence shows that Hispanic and Black borrowers have less access to credits or pay a higher price for mortgage loans in the US  \citep{munnell1996mortgage, charles2008rates, hanson2016discrimination, bayer2018what}. Outside the US, gender disparities are a major concern of discrimination. \cite{alesina2013do} find that Italian women pay more for overdraft facilities than men. \cite{bellucci2010does} and \cite{andres2021the} show that female entrepreneurs face tighter credit availability in Italy and Spain. \cite{ongena2016gender} document a strong correlation between gender bias and credit access across developing countries. In the modern world of Fin-Tech markets, \cite{bartlett2022consumer} shows that algorithmic lending reduces rate disparities between Latinx/African-American borrowers and other borrowers in consumer-lending markets but cannot eliminate the bias. \cite{fuster2022predictably} find that, in the US mortgage market, Black and Hispanic borrowers are disproportionately less likely to gain from the introduction of machine learning in lending decisions. 

Central in the welfare judgment of algorithmic lending is the trade-off between efficiency (controlling default risk) and equality (non-disparate treatment). The challenge of coping with this trade-off arises in part from the nature of discrimination, whether the discrimination is taste-based or statistical. If discrimination is primarily due to individual tastes that are unrelated to productivity, imposing the non-disparity regulation will not lead to strong tension between fairness and efficiency. However, if statistical discrimination dominates, i.e., the observed social feature being discriminated against is correlated with the unobserved productive feature, obeying the non-disparity regulation may hurt efficiency, for instance, granting credits to excessively risky borrowers. In practice, taste-based and statistical discriminations are hard to separate, making lenders struggle in an uncertain decision-making situation. This struggle is intensified by the tradeoff between type I and type II errors, whose consequences depend on the lender’s ability to assess and control risks.

In this section, we illustrate how our proposed algorithms help address the above challenge in finance management, using an example of potential gender bias in credit card consumption in Taiwan. The Taiwanese credit card dataset is from \cite{yeh2009comparisons}, which has been widely used to evaluate various data mining techniques. It is simple and transparent and has clear labeling of payment status that enables an analysis of financial risk.

This dataset contains information on the granted credit, demographic features, and payment history of $30{,}000$ individuals from April 2005 to September 2005. Importantly, it includes a binary status of the payment: either default, encoded by $0$, or non-default, encoded by $1$. Among all $30{,}000$ records, $6{,}636$ of them are labelled as $0$, i.e., default. The payment status defines the type I/II errors in the classification problem, and the protected attribute is gender. In the dataset, $11{,}888$ people are labeled as male and $18{,}112$ as female. Fitting such a typical credit-lending problem into the NP-EO classification framework, lenders (the Taiwanese banks) primarily want to control the risk of misclassifying someone who will default as non-default (type I error), although they also desire to minimize the chance of letting go of non-defaulters (type II error). Furthermore, by regulation or as a social norm, banks are not allowed to discriminate against qualified applicants based on gender. Therefore, to obtain the dual goal of risk control and fairness, our classification problem must satisfy both the NP and EO constraints.  Since we already illustrated in  Supplementary Material F.3 that the NP classifier mixed with random guesses performs worse than our proposed algorithms in all simulation settings, we do not include it in this real data section.

We use $1/3$ of the data for training and the other $2/3$ for testing, with stratification in both the protected attribute and the label. As an illustrative example, we set $\alpha = 0.1$, $\delta = 0.1$, $\varepsilon = 0.05$ and $\gamma = 0.1$. 
The base algorithm used is random forest. The process is repeated $1{,}000$ times, and the numerical results are presented in Table \ref{tbl:creditcard}. Using the classical classifier, the high-probability EO constraint is satisfied. Indeed, the EO violation rate is $0$, indicating that the random forest under the classic paradigm is ``fair'' and ``equal'' in terms of gender. This is not surprising, given that gender bias in modern Taiwan is not a significant concern. The problem with this classifier is that it produces a type I error of $0.633$, prohibitively high for most financial institutions. Benchmarked against the modest NP constraint ($\alpha = 0.1$), the violation rate is $1$,  
imposing high risk to the banks.

\begin{table}
\caption{\footnotesize Averages of type I/II errors, and type II error disparities, along with violation rates of NP and EO constraints over 1,000 repetitions for credit card dataset. Standard error of the means ($\times 10^{-4}$) in parentheses.}\label{tbl:creditcard}
\centering
\fbox{%
\begin{tabular}{*{6}{c}}
 & \begin{tabular}[c]{@{}l@{}}average of \\ type I errors\end{tabular} & \begin{tabular}[c]{@{}l@{}}average of \\ type II errors\end{tabular} & \begin{tabular}[c]{@{}l@{}}average of \\ type II error \\ disparities \end{tabular} & \begin{tabular}[c]{@{}l@{}}NP violation \\ rate\end{tabular} & \begin{tabular}[c]{@{}l@{}}EO violation \\ rate\end{tabular} \\ \hline
$\text{NP-EO}_{\text{OP}}$ & $.081(3.1)$ & $.719(6.8)$ & $.022(4.4)$ & .$039(61.3)$ & $.027(51.3)$ \\ 
$\text{NP-EO}_{\text{MP}}$ & $.088(3.1)$ & $.701(6.4)$ & $.025(4.6)$ & $.117(101.7)$ & $.047(67.0)$ \\ 
NP & $.088(3.1)$ & $.701(6.4)$ & $.050(4.1)$ & $.116(101.3)$ & $.456(157.6)$ \\
classical & $.633(3.9)$ & $.058(1.3)$ & $.017(1.4)$ & $1(0)$ & $0(0)$ \\
\end{tabular}}
\end{table}

When the NP paradigm alone is employed, the EO violation rate surges to $0.456$,  demonstrating a conflict between the banks’ private gain of improving risk control and the society’s loss of achieving fairness. 
When the $\text{NP-EO}_{\text{OP}}$ and $\text{NP-EO}_{\text{MP}}$ algorithms are employed, both the NP and EO constraints are satisfied with very small violation rates, and the classifiers simultaneously achieve the goals of risk control and fairness. The cost that the banks have to bear is missing some potential business opportunities from non-defaulters, which is reflected in the higher overall type II error committed by $\text{NP-EO}_{\text{OP}}$ algorithm. Consistent with the simulation results in Section \ref{sec:simulation}, compared to $\text{NP-EO}_{\text{OP}}$, $\text{NP-EO}_{\text{MP}}$ produces a smaller overall type II error while maintaining satisfactory (yet larger) violation rates.

The above example demonstrates the advantages and limitations of our proposed method in handling a real-world situation where the tradeoff between type I and type II errors is substantial, and a social constraint potentially exacerbates this tradeoff. The applicability of our approach hinges on decision makers’ assessment of the source of discrimination and the value of their targeted clients. For instance, a mature financial institution that worries more about risk control would not mind letting go of many new business opportunities.

\section{Discussion}\label{sec:discussion}

This paper is motivated by two practical needs in algorithmic design: a private user’s need to internalize social consideration and a social planner’s need to facilitate private users’ compliance with regulation. The challenge in fulfilling these needs stems from the conflict between the private and social goals. Notably, the social planner’s promotion of fairness and equality may constrain private users’ pursuit of profits and efficiency. In an ideal world without measurement and sampling problems, such a private-public conflict can be best resolved by maximizing a social welfare function with well-defined private and public components, and statistical tools hardly play any role. However, when knowledge about the social welfare function is partial, measurement of each component in the objective is imperfect, and consequences of predictive errors are uncertain, statistical innovation is called to resolve the private-public conflict. Our work is a response to this challenge.

We do not claim that our proposed NP-EO paradigm is superior to other classification paradigms. Rather, we propose an alternative framework to handle private vs. social conflicts in algorithmic design. Central in our analysis is the perspective of gaining security through statistical control when multiple objectives have to be compromised. The key to our methodological innovation is a principled way to redistribute specific errors so that the resulting classifiers have high probability statistical guarantees. Such finite-sample-based high probability guarantees have been the objectives of quite some previous work on algorithmic fairness, such as in \cite{Romano_2020}, \cite{Rava_fair}, and \cite{Linjun_Fair_2023}.

 Possible future research directions include but are not limited to (i) extending the solutions to multiple constraints concerning the social norms, which can involve multiple attributes such as race and gender, or multiple levels for one sensitive attribute such as race, (ii) working with parametric models, such as the linear discriminant analysis (LDA) model, to derive model-specific NP-EO classifiers that address small sample size problem and satisfy oracle type inequalities, (iii) replacing type I error constraint by other efficiency constraints, or including multiple fairness metrics, such as Equalized Odds, and (iv) studying fairness under other asymmetric efficiency frameworks such as isotonic subgroup selection in \cite{muller2024isotonic}.     

\includepdf[pages=-]{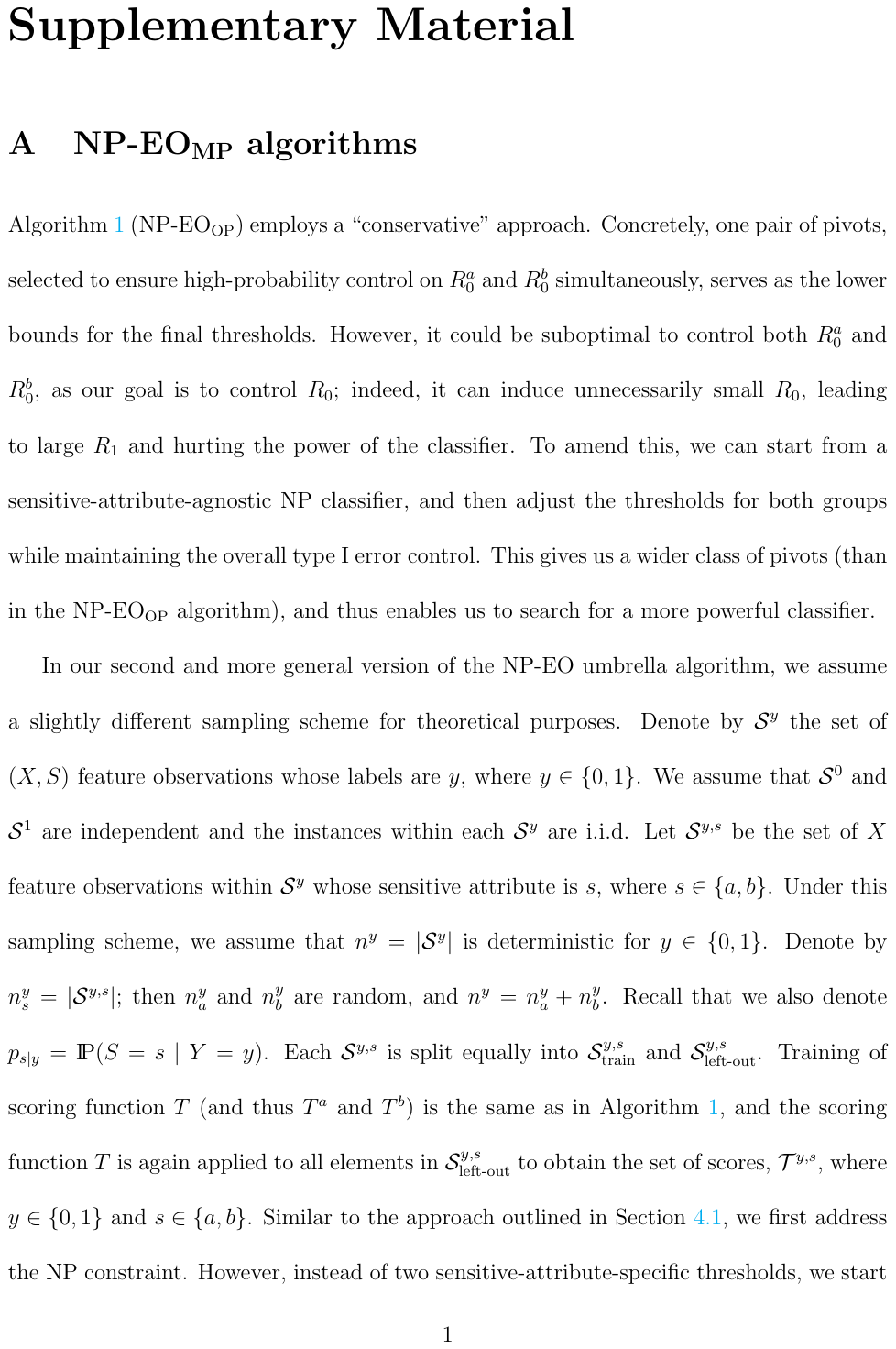}
\bibliography{fairness} 
\bibliographystyle{unsrtnat}

\end{document}